\documentclass[12pt]{iopart}

\newcommand{\re}{\ref}
\newcommand{\be}{\begin{equation}}
\newcommand{\ee}{\end{equation}}
\newcommand{\la}{\label}
\newcommand{\ber}{\begin{eqnarray}}
\newcommand{\eer}{\end{eqnarray}}

\newcommand{\bra}{\langle}
\newcommand{\ket}{\rangle}
\usepackage{iopams}  
\usepackage{graphicx}
\usepackage{epsfig}
\begin{document}
\maketitle
\eqnobysec

\title{
How to increase the applicability of integral transform approaches in physics?}
%How to extend the application of integral transform approaches to physical problems}

\author{G Orlandini$^1$, W Leidemann$^1$, V D Efros$^2$, N Barnea$^{3}$}

\address{$^1$ Dipartimento di Fisica, Universit\`a di Trento 
and Istituto Nazionale di Fisica Nucleare, Gruppo Collegato di Trento,  
I--38100 Trento Italy}

\address{$^2$ Russian Research Centre `Kurchatov Institute',  
Kurchatov Square, 1, 123182 Moscow, Russia }

\address{$^3$ The Racah Institute of Physics, The Hebrew University,   
91904 Jerusalem, Israel}

\ead{orlandin@science.unitn.it}

\date{today}
\begin{abstract}
 
Integral transform approaches  are numerous in many fields of physics, but in most
cases limited to the use of the Laplace kernel. However,
it is well known that the inversion of the Laplace transform is very problematic, so that
the function related to the physical observable is in most cases unaccessible. 
The great advantage of kernels of bell-shaped form has been demonstrated in 
few-body nuclear systems. In fact the use of the Lorentz kernel has allowed to
overcome the stumbling block of the ab initio description of 
reactions to the full continuum of systems of more than three particles. 
The problem of finding kernels of similar form, applicable to many-body problems 
deserves particular attention. If this search were successful the integral transform 
approach might represent the only viable ab initio access 
to many  observables that are not calculable directly.

\end{abstract}

%\tableofcontents

%*********************************************************
%*********************************************************
\section{Introduction}\label{sec:INTRO}
%*********************************************************
%*********************************************************

Integral transform approaches to physical problems are common to many fields in physics, e.g.
nuclear physics,  condensed matter physics and quantum chromo-dynamics. 
The motivation to use such approaches originates from the fact that,
while one is often unable to evaluate a physical quantity of interest $\cal X$, directly, one can calculate a few 
of its integral properties. Particularly interesting are integral transforms.
% From these one can infer 
They can give  information on $\cal X$. However,
a sufficiently detailed reconstruction of $\cal X$  (inversion of the transform) is sometimes a formidable task 
or even impossible. 
One of the reasons is that in many cases the integral property known to be accessible by a numerical technique
is the Laplace transform ($\cal LT$). This occurs in the numerous physical processes that are 
described by diffusion equations. Problems involving $\cal LT$s also arise in computing statistical functions, which
is very common in many fields.

From the Mellin's inverse formula (Bromwich integral) for the inverse Laplace transform~\cite{DAVIES:2002}, 
one  sees that it is necessary that the $\cal LT$  is computable in the complex half-plane of convergence. 
In this case $\cal X$  can be computed by evaluation
of a complex  path integral, which can be a formidable task.
However, in the common situation when the Laplace transform is computable (or measurable) on the real 
and positive axis only, the problem is extremely {\it ill-posed}. This case is much more complicated 
because of the absence of an exact inversion formula. In addition, as it will be explained below, one can say that
among  {\it ill-posed} problems, the inversion of the $\cal LT$ is one of the worst. 
There are some more or less efficient numerical methods available for computing the inverse 
transform~\cite{JAYNES:1978,KRYZHNIY:2004}, however, in general 
when the input is numerically noisy and incomplete it may be impossible to obtain a stable result~\cite{ACTON:1970}.

In the last fifteen years it has been shown that an integral transform approach with a {\it bell-shaped}
kernel, namely  the Lorentzian function~\cite{EFROS:1994},  
helps in solving a longstanding problem in few-body physics, i.e. 
the calculation of reactions involving states in the far continuum of four- (\cite{EFROS:1997a}-\cite{BACCA:2009})
and even six-~\cite{BACCA:2002} and seven-~\cite{BACCA:2004a} body systems
(for a review see~\cite{EFROS:2007}). The scope of this contribution is to (i) clarify  
why the Lorentzian function is a {\it good} kernel
(section 1), (ii) discuss the limits of this approach (section 2) (iii) define the {\bf open problems} (section 3) 
and (iv) suggest possible directions to explore for finding solutions (section 4).  

\section{Why is the Lorentzian function is a {\it good} kernel ?}\label{good}

\begin{figure}[htb] %1
\bigskip\bigskip\bigskip
\centering
\includegraphics[angle=0,width=.80\textwidth]{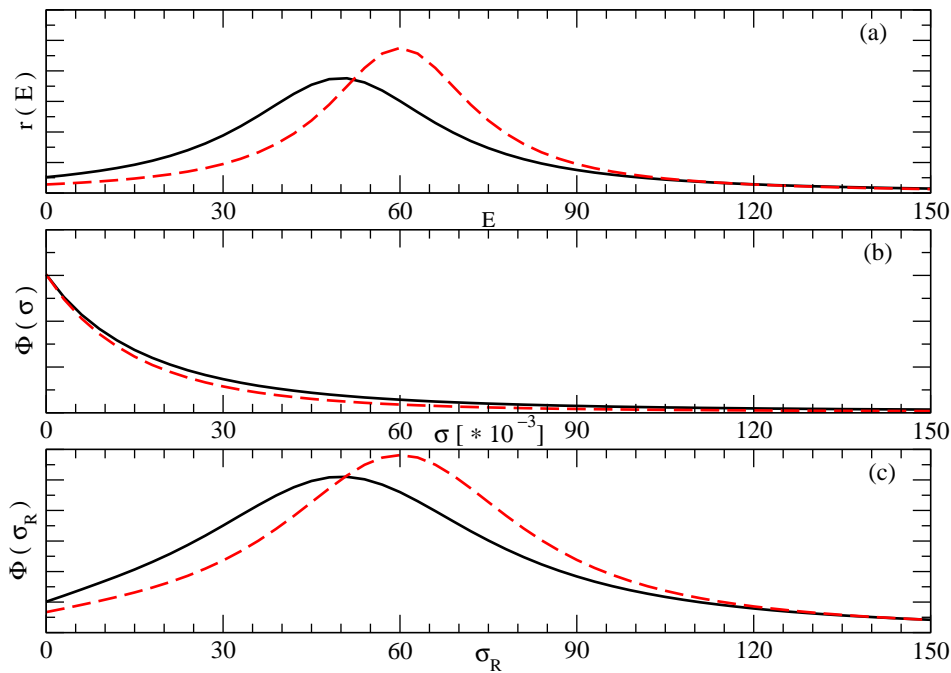}
\caption{(a): The functions $r_1$ and $r_2$, (b): their Laplace transforms, (c) their Lorentz transforms with $\sigma_I=10$.}
\label{fig1}
\end{figure}

To illustrate what is the intrinsic origin of the difficulties in inverting transforms with the Laplace kernel
or, similarly, with the Stieltjes kernel~\cite{EFROS:1985,EFROS:1993},  consider
figure~\ref{fig1}. The upper panel shows two  functions $r_1(E)$ and $r_2(E)$, 
while in the middle panel their $\cal LT$s are plotted, i.e. 
\begin{equation}
\Phi(\sigma)=\int\, K(\sigma,E)\,r(E)\,dE\,,\la{phi}
\end{equation}
with $K(\sigma,E)=e^{-\sigma E}$. As one can see the curves in figure~\ref{fig1}(a) do not resemble the curves
in figure~\ref{fig1}(b).
The functios $r_i(E)$ exhibit a peak, while the $\cal LT$s are monotonically decreasing. 
Moreover, one sees that though $r_1(E)$ and $r_2(E)$ are rather different they lead to very similar $\cal LT$s.
Considering that both transforms could lie within the error of a numerical calculation 
one understands that it is impossible by inversion to distinguish between 
the curves in~\ref{fig1}(a). The reason is that the Laplace kernel has spread the information 
over a very large domain. 
As shown in~\ref{fig1}(c) the situation is different for the Lorentz integral transform (LIT), where the kernel of
the transform is the Lorentzian function 
\begin{equation}
  K(\sigma_R,\sigma_I,E)= \frac{1}{(E-\sigma_R)^2+\sigma_I^2}\,.
\end{equation}

The form of the functions in~\ref{fig1}(c) is preserved and the relative difference between the LITs, 
is of the same order as it is for the functions in~\ref{fig1}(a). 

We do not intend to discuss here the problem of the inversion of integral transforms, which are known in mathematics
under the name of {\it ill-posed problems}. An extended literature exists about this topic (see e.g.~\cite{TIKHONOV:1977}). 
We  only want to mention that, contrary to what the unfortunate name might suggest, 
they can be solved using regularization methods. Such methods  are very efficient when the kernel
is bell shaped with a controllable width. A clarifying discussion about this point can be found 
in~\cite{BARNEA:2010}. 

In relation to what has been said above, one could think that the perfect kernel is a $\delta-$function, i.e. 
$K(\sigma_R,E)=\delta(\sigma_R-E)$.
In fact in this case the function and its transform would coincide. While this 
clearly does not help, the Lorentzian function, which up to normalization turns to the $\delta$-function at
$\sigma_I\rightarrow 0$,
has the necessary characteristics, as have the Gaussian functions or other bell-shaped kernels.
 
However, it is clear that the other {\it conditio sine qua non} for applying an integral transform approach
to physical problems is that the transform is calculable in practice. 
In~\cite{EFROS:1994} it was suggested that the Lorentzian kernel might satisfy both requisites when applied 
to a specially difficult and unsolved problem, i.e. the {\it exact} (ab initio) calculation  
of perturbation induced reactions to  states in the continuum (scattering states) of a many-body system.
In the following we will shortly summarize why this is the case. More extensive information can be found 
in~\cite{EFROS:2007}, where the successful applications of what is now known as the Lorentz integral transform  
method have been reviewed and the procedure to apply it also to non-perturbative reactions has been outlined.

\subsection{Short summary of the LIT method}

Very often physical quantities of interest have (or may be reformulated to have) the following structure
\begin{equation}
r(E)=
\sum\!\!\!\!\!\!\!\int\,d\gamma
\langle Q|\Psi_\gamma\rangle\langle\Psi_\gamma|Q'\rangle
\delta(E_\gamma-E), \label{rr}
\end{equation}
where the integration and summation  go 
over all the complete and orthonormal set of eigenstates $|\Psi_\gamma\rangle$ of an hamiltonian $\hat H$
and the norms $\langle Q|Q\rangle$ and $\langle Q'|Q'\rangle$ are finite.

When the energy $E$ and the number of particles in a system increases
the direct calculation of the quantity $r(E)$
becomes prohibitive. The difficulty is related to the fact that in these cases
a great number of continuum spectrum
states $|\Psi_\gamma\rangle$ contribute to $r(E)$ and the structure of these  states is very complicated. 

In order to overcome the problem one can  consider instead 
an integral transform
%\be
%\Phi(\sigma)=\sum\!\!\!\!\!\!\!\int\, K(\sigma,E)\,r(E)\,dE\la{phi}
%\ee
with a smooth kernel $K(\sigma,E)$. This yields
\begin{eqnarray}
\Phi(\sigma)&=&\sum\!\!\!\!\!\!\!\int\,d\gamma
\langle Q|\Psi_\gamma\rangle K(\sigma,E_\gamma)
\langle\Psi_\gamma|Q'\rangle\nonumber\\
&=&\sum\!\!\!\!\!\!\!\int\,d\gamma
\langle Q|\hat{K}(\sigma,\hat{H})|\Psi_\gamma\rangle 
\langle\Psi_\gamma|Q'\rangle.\la{r}
\end{eqnarray}
Using the closure property one obtains
\be
\Phi(\sigma)=\langle Q|\hat{K}(\sigma,\hat{H})|Q'\rangle.\la{s}
\ee 

The right--hand side of~(\re{s})  requires only the knowledge of the finite norm functions 
$|Q\rangle$ and $|Q'\rangle$ and avoids completely the full knowledge 
of the spectrum of $\hat H$. (It is clear that what is presented here
can be considered as a generalization of the sum rule, namely the {\it moment method}).

Choosing for  $K(\sigma,E)$ the Lorentzian function 
\begin{equation}
K(\sigma,E)=\frac{1}{(E-E_0-\sigma_R)^2+\sigma_I^2}\,.\la{kRI}
\end{equation}
where
\be
\sigma=E_0+\sigma_R+i\sigma_I\,\,\,\,\,\,\,\,\,\,\,\,\sigma_I\neq0,\la{not}\,
\ee
and $E_0$ is the ground--state energy,  
one has
\begin{eqnarray}\label{L_ssp}
L(\sigma_R,\sigma_I)
&=&\bra Q | \frac{1}{(\hat{H}-E_0-\sigma_R+i\sigma_I )}
           \frac{1}{(\hat{H}-E_0-\sigma_R-i\sigma_I )}| Q'\ket\nonumber\\
&\equiv&\langle\tilde\Psi|\tilde\Psi'\rangle\,, \label{ll}
\end{eqnarray}
where the `LIT functions' $\tilde{\Psi}$ and $\tilde{\Psi}'$ are solutions to the inhomogeneous equations
\begin{eqnarray}
\left(\hat{H}-E_0-\sigma_R-i\sigma_I\right)|\tilde{\Psi}\rangle&=&|Q\rangle\,, \la{eq1}\\
\left(\hat{H}-E_0-\sigma_R-i\sigma_I\right)|\tilde{\Psi}'\rangle&=&|Q'\rangle\,. \la{eq2}
\end{eqnarray}
Let us  suppose that  $|Q\rangle=|Q'\rangle$). In this case  $L(\sigma)$
equals to $\langle\tilde{\Psi}|\tilde{\Psi}\rangle$ ($\langle\tilde{\Psi}'|\tilde{\Psi}'\rangle$). 
Since for $\sigma_I\ne0$
the integral in (\re{phi}) does exist, the norm of $|\tilde{\Psi}\rangle$ ($|\tilde{\Psi}'\rangle$)
is finite. This implies
that $|\tilde{\Psi}\rangle$ and $|\tilde{\Psi}'\rangle$ are  {\em localized} functions.
%Consequently,~(\re{eq1}) and~(\re{eq2}) can be solved with bound--state type methods.
%Similar to the problem 
%of calculating a bound state it is sufficient to impose the only condition that the 
%solutions  of~(\ref{eq1}) and~(\ref{eq2}) are localized. 
This means that
in contrast to continuum spectrum problems, in order 
to construct a solution, it is not necessary here to reproduce a complicated large 
distance asymptotic behaviour in the coordinate representation or singularity structure 
in the momentum representation. This is a very substantial simplification.

\section{What are the present limits of LIT applications ?}\label{limits}  

The LIT method has been successfully applied to quite a few physical problems.
They regard electroweak
reactions with {\bf few-body} systems. An extension to  {\bf many body} systems, however, encounters at present 
some limits. They become clear in the following where we describe
how one calculates the LIT $L(\sigma_R,\sigma_I)$ of equation~(\ref{ll})..

Most of the LIT applications have been solved by using complete many-body basis and 
expanding $|\tilde{\Psi}\rangle$ and $|\tilde{\Psi}'\rangle$  
over $N$ localized states. 
A convenient choice is to use $N$ linear combinations of states that diagonalize the hamiltonian matrix. 
We denote these combinations $|\varphi_\nu^N\rangle$ and the eigenvalues $\epsilon_\nu^N$. 
The index $N$ is to remind that they  both depend on $N$. 
If the continuum starts at $E = E_{th}$ then at sufficiently high $N$ the states $|\varphi_N\rangle$ having 
$\epsilon_\nu^N< E_{th}$ will represent approximately the bound states. The other states  will 
gradually fill in the continuum as $N$ increases. The expansions of our localized  LIT functions 
read as 
\begin{equation}\label{nu1}
|\tilde{\Psi}\rangle =\sum_\nu^N\frac{\langle\varphi_\nu^N|Q\rangle}
{\epsilon_\nu^N-E_0-\sigma_R-i\sigma_I}|\varphi_\nu^N\rangle\,,
\end{equation}
\begin{equation}\label{nu2}
|\tilde{\Psi}'\rangle=\sum_\nu\frac{\langle\varphi_\nu^N|Q'\rangle}
{\epsilon_\nu^N-E_0-\sigma_R-i\sigma_I}|\varphi_\nu^N\rangle\,.
\end{equation}	
Substituting  (\ref{nu1}) and (\ref{nu2}) into (\ref{L_ssp}) 
yields the following expression for the LIT,
\begin{equation}\label{L_epsnu}
L(\sigma_R,\sigma_I)=\sum_{\nu} \frac{\bra Q| \varphi_\nu^N \ket \bra \varphi_\nu^N | Q' \ket}
                    {(\epsilon_{\nu}^N-E_0-\sigma_R)^2+\sigma_I^2}  \;,
\end{equation}
and for the common case $|Q '\ket=|Q \ket$
\begin{equation}\label{L_epsnu_diag}
L(\sigma_R,\sigma_I )=\sum_{\nu} \frac{ |\bra \varphi_\nu^N | Q \ket|^2}
                     {(\epsilon_{\nu}^N-E_0-\sigma_R)^2+\sigma_I^2}\;.\label{gg}
\end{equation}
From~(\ref{L_epsnu}) and (\ref{L_epsnu_diag}) it is clear that in this way
$L(\sigma_R,\sigma_I)$ becomes a sum of Lorentzians. The spacing between 
the corresponding eigenvalues with $\epsilon_\nu^N > E_{th}$ depends on $N$ and in a given energy 
interval the density of these eigenvalues increases with $N$. 
(Since the extension of the basis states grows with $N$, this resembles the increase 
of the density of states in a box, when its size increases). 
For the reliability of the inversion  one needs to reach the regime where one has 
a sufficient number of levels $\epsilon_\nu^N$ in a $[\sigma_R - \sigma_I, \sigma_R + \sigma_I]$ interval.

In a similar way one can obtain other integral transforms with bell shaped kernels e.g. the Gauss integral transform (GIT)
$K(\sigma_R,\sigma_I, E)= e^{-(E-\sigma_R)^2/\sigma_I^2}$: 

\begin{equation}\label{L_epsnu_diag}
G(\sigma_R,\sigma_I )=\sum_{\nu}|\bra \varphi_\nu^N | Q \ket|^2 e^{-(\epsilon_{\nu}^N-\sigma_R)^2/\sigma_I^2}
                    \;.\label{gg}
\end{equation}

It is clear that for a many-body system this is a formidable task, because of the larger and larger degeneracy
of states with increasing N and the little chance that out of the many states a significant number is found within
the $\sigma_I$ extension. 
In this resides the limit of the LIT method if one wants to extend it to many-body systems. 

A more efficient  procedure can be applied using the Lanczos approach.
This  consists in rewriting $L(\sigma_R,\sigma_I)$
as 
\begin{equation} \label{lorelan1}
 L(\sigma)=-\frac{1}{\sigma_I} \mbox{Im} 
\left\{ \bra Q | \frac{1}{\sigma_R+i\sigma_I+E_0-
\hat{H}} | Q'\ket \right\}\:\mbox{.}
\end{equation}
and then as a continuous fraction in function of the Lanczos coefficients $a_i$ and $b_i$
\begin{equation} \label{lorelan3}
L(\sigma )=-\frac{1}{\sigma_{I}}\mbox{Im}\left \{ \frac
          {\bra Q | Q' \ket
           }{z-a_{0}-\frac{b^{2}_{1}}{z-a_{1}-\frac{b^{2}_{2}}{
          z-a_{2}-b^{2}_{3}\ldots}}}\right \} \:.
\end{equation}
This has the big advantage that one has not to diagonalize or invert the hamiltonian matrix,
however the problem of extending the method to many-body systems remains similar to the previous case.

{\bf Important remarks:} 
\begin{itemize}
\item We would like to emphasize that, contrary to discretization approaches to the continuum,
here the use of a discrete basis is perfectly legitimate, because of
the bound state like boundary conditions of $|\tilde\Psi\rangle (|\tilde\Psi'\rangle)$.
The continuum spectrum is recovered with a controlled resolution by the regularization when inverting the transform. 
\item One can notice that equation~(\ref{gg}) corresponds to the following common procedure.
This consists in diagonalizing the hamiltonian on a discrete basis and then  folding 
the $|\bra \varphi_\nu^N | Q \ket|^2$ with a Lorentzian function (or a Gaussian function), in order to simulate 
the ``experimental window``.
However, as equation (\ref{gg}) clearly states, this {\bf does not} represent $r(E)$ itself, but its LIT (GIT),
that has to be inverted (see~\cite{EFROS:2007}, figure 5 and relative discussion).
%\item In practice, if by a diagonalization procedure one knows $|\bra \varphi_\nu^N | Q \ket|^2$,  folding it
%with any bell shaped $K(\epsilon_\nu^N,\sigma)$ gives the corresponding integral transform of $r(E)$ with the Kernel
%$K(E,\sigma)$. The inversion of the different transforms obtained with different kernels should converge 
%to the same stable $r(E)$.
\item Considering our discussion one realizes that,  within the eigenvalue approach, the difficulties in generalizing
the integral transform method to many body systems do not lie in the  kernels, that may be chosen  conveniently, 
but in handling very large basis.
\end{itemize}

\section{Open problems}

The problem of calculating exactly (ab initio) reactions to continuum states (scattering states) of many-body systems
is one of the oldest  open problems in Quantum Mechanics. Many important steps have been made in the last years 
towards a solution, thanks to the integral transform method. However, much has still to be done and we summarize 
the main problems in the following.
\begin{itemize}
\item {\bf Application of the LIT method to few-body non-perturbative processes} 
(hadron  scattering). The LIT method has been applied successfully only to inclusive and exclusive 
perturbation induced processes. The formalism for applications to non perturbative processes already exists. 
One can find it briefly discussed in
section 2.3 of~\cite{EFROS:2007}. It is essentially  a reformulation with the Lorentz kernel 
of the original idea in~\cite{EFROS:1985} with the Stieltjes kernel. 
\item {\bf Spectra presenting narrow resonances.} From (\ref{eq1}) and (\ref{eq2}) it is clear that 
the smaller the width $\sigma_I$
the more slowly the LIT functions $\tilde\Psi$ approach zero at large distances, making it very 
hard to find accurate transforms~\cite{LEIDEMANN:2008}. In particular using expansions of $|\tilde \Psi\rangle $
over basis states one has to make sure that a sufficient  number of $|\phi_\nu^N\rangle$ states 
have energies $\epsilon_\nu^N$ in the resonance region
Therefore very narrow resonances can be hardly resolved. 
\item {\bf Integral transforms and many-body methods.} In section~\ref{limits} it has been explained
that for an extension of the LIT method to a larger number of particles one has to be able to handle very large basis.
Investigating the use of the Coupled Cluster method to calculate the LIT is one of the interesting 
directions~\cite{BACCA:2010}.
In alternative one could search for other  kernels, that allow 
the calculation of the transform by some other methods. Due to the large use of MC approaches, 
not only in nuclear physics, but also in many other fields, from condensed matter to QCD, it would be very interesting
to explore the applicability of Monte Carlo (MC) techniques to integral transforms with kernels other than the Laplace one. 
In the following section we suggest possible investigations along this line. 
\end{itemize}

\section{New directions to explore}

As discussed in section~\ref{good} a good kernel has to be bell-shaped and has to generate
a transform that can be calculated. 
The Laplace kernel leads to transforms of the kind
\be
\Phi(\tau)=\langle Q|e^{[-(\hat H-E_0)\tau]}|Q'\rangle.\label{IT}
\ee 
that are usually calculated by MC techniques (evolution in
imaginary time). However, the kernel  does not have the desired form. 
One difference between the Laplace and the Lorentz kernel is that the former is a function of one parameter 
(besides the integration variable), while the latter is a function of two parameters that have a specific function:
one allows to regulate the width of the bell, the other, is connected to its maximum  and allows {\it to scan} the function
of interest over its domain.
Therefore one should think of other two or more parameter kernels that are such combinations of exponentials to 
generate bell shaped kernels.
One of them could be e.g.~\cite{GATTI:2009}

\begin{equation}
K(\sigma,\tau_1,\tau_2,E)= \Theta(E-\sigma)\left[e^{[-(E-\sigma)\,\tau_1]}- 
e^{[-(E-\sigma)\,\tau_2]}\right]\,\,\,\,\,\tau_2>\tau_1 > 0\,.
\end{equation}
This is a bell shaped function that, with suitable choices of the parameters,  allows to vary at wish both the width
and the centroid. The Heaviside step function $\Theta$  ensures both that the exponents remain negative and that 
$K(\sigma,\tau_1,\tau_2,E)$ is positive definite.
What is to investigate is whether a MC algorithm could be deviced to calculate 
\be
\Phi(\sigma,\tau_1,\tau_2,)=\langle Q|\Theta(\hat H-\sigma)\left[e^{[-(\hat H-\sigma)\,\tau_1]}- 
e^{[-(\hat H-\sigma)\,\tau_2]}\right]|Q'\rangle\,,
\ee 
via imaginary time evolution as (\ref{IT}).
Alternatively one can think of another representation of the $\delta$-function, i.e.
\begin{equation}
K(\sigma,\tau,E)=\tau e^{-|(E-\sigma)|\, \tau}\,
\end{equation}
where again $\tau$ and $\sigma$ allow to change width and centroid of the kernel, respectively.
The presence of the absolute value in the exponent again requires to deal with operators that contain 
the $\Theta$ operator. This might represent a problem.  

Given the large number of physical problems in many different fields, that are connected to Laplace transforms,
an effort to investigate the use of kernels like those proposed here, if successful, could open the way to
many very interesting developments.

\end{document}